\documentclass[twocolumn,showpacs,amsmath,nofootinbib,pra,aps,amssymb,longbibliography]{revtex4-2} %
\usepackage[T1]{fontenc}
\usepackage[utf8]{inputenc}
\usepackage{times}
\usepackage{color} 
\usepackage{array}
\usepackage{amssymb,amsmath}
\usepackage{amsfonts}
\usepackage{amsbsy}
\usepackage{wasysym}
\usepackage[pdftex]{graphicx} 
\usepackage{bm}
\usepackage{float}
\usepackage{dcolumn}
\usepackage{textcomp}
\usepackage{scalerel}
\usepackage{tikz}
\usetikzlibrary{svg.path}

\definecolor{orcidlogocol}{HTML}{A6CE39}
\tikzset{
  orcidlogo/.pic={
    \fill[orcidlogocol] svg{M256,128c0,70.7-57.3,128-128,128C57.3,256,0,198.7,0,128C0,57.3,57.3,0,128,0C198.7,0,256,57.3,256,128z};
    \fill[white] svg{M86.3,186.2H70.9V79.1h15.4v48.4V186.2z}
                 svg{M108.9,79.1h41.6c39.6,0,57,28.3,57,53.6c0,27.5-21.5,53.6-56.8,53.6h-41.8V79.1z M124.3,172.4h24.5c34.9,0,42.9-26.5,42.9-39.7c0-21.5-13.7-39.7-43.7-39.7h-23.7V172.4z}
                 svg{M88.7,56.8c0,5.5-4.5,10.1-10.1,10.1c-5.6,0-10.1-4.6-10.1-10.1c0-5.6,4.5-10.1,10.1-10.1C84.2,46.7,88.7,51.3,88.7,56.8z};
  }
}

\newcommand\orcidicon[1]{\href{https://orcid.org/#1}{\mbox{\scalerel*{
\begin{tikzpicture}[yscale=-1,transform shape]
\pic{orcidlogo};
\end{tikzpicture}
}{|}}}}

\usepackage[unicode,breaklinks]{hyperref}
\hypersetup{
    unicode=true,
    plainpages=false, 
    colorlinks=true,
    linkcolor=blue,
    citecolor=blue,
    filecolor=black,
    urlcolor=blue
}
\newcommand{\arot}{a_\mathrm{rot}^*}

\newcommand{\gdd}{g_\mathrm{dd}}
\newcommand{\add}{a_\mathrm{dd}}
\newcommand{\edd}{\epsilon_\mathrm{dd}}
\newcommand{\br}{\mathbf{r}}
\newcommand{\bx}{\mathbf{x}}

\newcommand{\bk}{\mathbf{k}}
\newcommand{\UDt}{\tilde{U}_\mathrm{dd}}
\newcommand{\Phidd}{\Phi_\mathrm{dd}}
\newcommand{\UD}{U_\mathrm{dd}}
\newcommand\gammaQF{\gamma_\mathrm{QF}}

\newcommand\brho{\boldsymbol{\rho}}

\newcommand{\LGP}{\mathcal{L}_\mathrm{GP}}
\newcommand{\ME}{\mathcal{E}}
\newcommand{\MF}{\mathcal{F}}
\newcommand{\MC}{\mathcal{C}}

\newcommand{\Ut}{\tilde{U}}

\newcommand\intuc{\int_{\mathrm{uc}}d \bx}
\DeclareMathOperator\Ei{Ei}
\renewcommand{\Re}{\operatorname{Re}}

\begin{document}

\title{Supersolidity and crystallization of a dipolar Bose gas in an infinite tube}
\author{Joseph~C.~Smith \orcidicon{0000-0002-2670-4044}}
\author{D.~Baillie \orcidicon{0000-0002-8194-7612}}
\author{P.~B.~Blakie \orcidicon{0000-0003-4772-6514}}

	\affiliation{Dodd-Walls Centre for Photonic and Quantum Technologies, New Zealand}
	\affiliation{Department of Physics, University of Otago, Dunedin 9016, New Zealand}
  
\date{\today}
\begin{abstract}
 We calculate the ground states of a dipolar Bose gas confined in an infinite tube potential. We use the extended Gross-Pitaevskii equation theory and present a novel numerical method to efficiently obtain solutions. A key feature of this method is an analytic result for a truncated dipole-dipole interaction potential that enables  the long-ranged interactions to be accurately evaluated within a unit cell. Our focus is on the transition of the ground state to a crystal driven by dipole-dipole interactions as the short ranged interaction strength is varied. We find that the transition is continuous or discontinuous depending upon average system density. These results give deeper insight into the supersolid phase transition observed in recent experiments, and validate the utility of the reduced three-dimensional theory developed in [Phys.~Rev.~Res.~{\bf 2}, 043318 (2020)] for making qualitatively accurate predictions.
\end{abstract}

\maketitle
\section{Introduction}

Supersolidity is a phase of matter that simultaneously exhibits crystalline order and superfluidity, and has a history dating back almost 70 years \cite{Gross1957a,Penrose1956a,Andreev1969a,Chester1970a,Gross1958a,Leggett1970a}. The solid phase of He\nobreakdash-4 was an early focus of work looking for supersolidity, but thus far it has not been found \cite{Greywall1997a,Kim2004a,Kim2004b,Kim2012a}. Over the past decade significant attention turned to ultra-cold atomic systems \cite{Boninsegni2012a} due to the high degree of experimental control over microscopic parameters.  In 2017 two groups reported supersolidity in optically-coupled Bose-Einstein condensates \cite{Leonard2017a,Li2017a}. This was followed by its realization with ultra-cold dipolar Bose gases \cite{Chomaz2019a,Bottcher2019a,Tanzi2019a}. In these experiments the system was initially cooled into a (superfluid) Bose-Einstein condensate, then the $s$-wave interaction was gradually reduced until the  magnetic dipole-dipole interactions (DDIs) dominated and caused the ground state to spontaneously modulate, i.e.~crystalline order emerged. In some situations the phase coherence was maintained across the system as the spatial modulation developed, indicating that a supersolid state was obtained.
Further evidence for this interpretation has been provided by the analysis of the low energy collective excitations \cite{Tanzi2019b,Guo2019a,Natale2019a}. These experiments  used cigar-shaped confinement potentials with the magnetic dipoles of the atoms polarized along a tightly confined direction. In this situation the modulation that develops is one-dimensional (1D)  and occurs along the most weakly confined direction (i.e~ the long-axis of the cigar-shaped trap).

The theoretical understanding of dipolar Bose gases is provided by the extended Gross-Pitaevskii equation (eGPE) theory \cite{Ferrier-Barbut2016a,Wachtler2016a,Bisset2016a}. This theory includes the leading order effects of quantum fluctuations \cite{Lee1957a,Lima2012a}, which are necessary to stabilize the condensate against mechanical collapse due to the attractive component of the DDIs. Applications of the eGPE theory to the experimental regime have provided a quantitative description of the experimental results and has played a key role in interpreting many experimental observations. However, these calculations are particular to the experimental system under consideration (i.e.~atom number and parameters of the three-dimensional (3D) harmonic confinement).

It is of interest to understand the crystallization and supersolid behavior of a dipolar Bose gas in an infinite tube system: a gas of average linear density $n$, confined by harmonic transverse confinement, but free to move along the tube. 
This case has a continuous translation symmetry along the tube which is broken if the system crystallizes, and constitutes the thermodynamic limit relevant to current experiments.  A related system was analysed by Roccuzzo \textit{et al.}~\cite{Roccuzzo2019a}, differing by the additional condition that the tube had fixed periodic boundary conditions (i.e.~a finite ring geometry). That work provided important insights to the transition for a particular choice of system density: they observed that as a function of the $s$-wave scattering length the superfluidity changed discontinuously as the transition to the crystalline (supersolid) state occurred. They also found that this transition was close to where a roton excitation softened to zero energy, marking the onset of a dynamical instability in the uniform superfluid state.
The infinite tube system (without periodic boundary conditions) was first examined by Blakie \textit{el al.}~\cite{Blakie2020b} employing a hybrid variational approach to simplify the 3D eGPE description to a 1D form (the reduced 3D theory). They considered a wide range of densities and produced a phase diagram showing that the nature of the crystalline transition depended on the system density: it was discontinuous in the low- and high-density regimes, but was continuous for an intermediate range of densities. Furthermore, where the crystallization was found to develop continuously, the transition point coincided precisely with the roton softening instability.
Predictions of the reduced 3D theory are known to have quantitative differences from full eGPE calculations (e.g.~see \cite{Blakie2020a}), which could affect the reliability of its transition predictions in the infinite tube system. This motivates a full numerical  study using the eGPE theory to clarify the crystallization transition in the infinite tube system.

In this work we report the first results of a full eGPE study of the infinite tube system. Calculations near the transition point must resolve very small energy differences between the modulated  and uniform states to predict the correct ground state, and hence determine if the transition is continuous. This demands accurate numerical calculations of the eGPE. The method we present utilizes a unit cell description of the eGPE, with the unit cell size being an optimized parameter. To calculate the singular and long-ranged DDIs accurately in this geometry we develop a new analytic result for a truncated DDI potential. Our results allow us to assess the phase diagram over a range of densities and make a comparison to the reduced 3D theory.

An outline of this paper is as follows.
In Sec.~\ref{Sec:Theory} we outline the eGPE theory and our numerical method for obtaining stationary states. As part of this we introduce a truncated DDI potential that allows us to accurately evaluate the DDIs in a unit cell of the infinite system. We also discuss measures of density modulation (crystallization) and superfluidity. Section~\ref{Sec:Results} contains our main results. We present a phase diagram over a wide range of densities encompassing regions where the transition is continuous and discontinuous. We consider examples of the continuous and discontinuous transitions in further detail, and verify that the roton softening coincides with the transition point when the transition is continuous.  The paper concludes in Sec.~\ref{Sec:Conlcusion}.

\section{Theory}\label{Sec:Theory}

\subsection{System}
We consider an ultra-dilute gas of highly magnetic bosonic atoms confined in a tube potential (transverse harmonic confinement) of the form  \begin{align}
V(\brho) = \tfrac12 m(\omega_{x}^{2}x^{2} + \omega_{y}^{2}y^{2}),
\end{align}
where $\omega_{x,y}$ are the angular trap frequencies, and $\brho=(x,y)$.
The eGPE energy functional for this system is given by $E=\int d\bx\,E(\bx)$ where 
\begin{align}
E(\bx) &= \psi^*[h_\mathrm{sp}+\tfrac12g_s|\psi|^2+\tfrac12\Phidd(\bx) +\tfrac25\gammaQF|\psi|^3]\psi,
\end{align}
and $h_\mathrm{sp}=-\frac{\hbar^2}{2m}\nabla^2+V(\brho)$ is the single particle Hamiltonian. The short ranged interactions are governed by the coupling constant $g_s= 4\pi \hbar^2 a_s/m$ where $a_s$ is the $s$-wave scattering length.
The long-ranged DDIs are described by the potential
\begin{align}
\Phidd(\bx)=\int d\bx'\,\UD(\bx-\bx')|\psi(\bx')|^2,\label{PhiDD}
\end{align}
where the atoms are polarized along $y$
\begin{align}
\UD(\br) = \frac{3\gdd}{4\pi r^3}\left(1-\frac{3y^2}{r^2}\right),\label{Udd}
\end{align}
and $\gdd=4\pi\hbar^2\add/m$, with $\add = m\mu_0\mu_m^2/12\pi\hbar^2$ being the dipolar length, and $\mu_m$ the magnetic moment. The effects of quantum fluctuations are described by the quintic nonlinearity with coefficient $\gammaQF = \frac{32}3 g_s\sqrt{a_s^3/\pi}\mathcal{Q}_5(\edd)$ where $\mathcal{Q}_5(x)=\Re\{\int_0^1 du[1+x(3u^2 - 1)]^{5/2}\}$ \cite{Lima2011a} and $\edd \equiv \add/a_{s}$. 

We consider the gas to be constrained to have an average linear density of $n$, and hence  the state $\psi$ is non-normalizable and the energy $E$ is infinite. For this reason it is necessary to minimize the energy per particle $\ME$ to determine the system ground state. We can categorise such ground states into two types. (i) \textit{Uniform states} that exhibit a continuous translational invariance and have the form $\psi(\bx)=\sqrt{n}\chi(\brho)$. (ii) \textit{Crystalline states} that break translational symmetry and vary periodically along $z$  with period $L$. Our main aim in this work is to understand where and how the system transitions between these two states as the linear density and short range interactions vary.

\subsection{Unit cell treatment}
It is useful to reduce the system to a unit cell (uc) of length $L$ along $z$ that periodically repeats along this direction, i.e.~we take the uc to be $z\in [-L/2,L/2)$, with $x$ and $y$ over all space, and $\psi(\bx)=\psi(\bx+L\hat{\mathbf{z}})$. The average density constraint enforces that $\intuc |\psi|^2=nL$ atoms are in the uc. 
The energy per particle is given by
\begin{align}
\ME=\frac{1}{nL}\intuc\, E(\bx).\label{ucEfunc}
\end{align}
For the case of a uniform state, $\ME$ is independent of $L$ (i.e.~the uc is arbitrary), whereas for a crystalline state a particular $L$ will minimise $\ME$ and the uc is well defined.\footnote{We note that due to periodicity in this case $\ME$ will also be an equivalent minimum for $2L,3L,\ldots$} We also note that the integration in (\ref{ucEfunc}) is restricted to the uc whereas the DDI term $\Phidd$ given by \eqref{PhiDD} involves an integral over all space.

\subsection{Numerical method overview} 

For a fixed uc the energy of the system can be reduced by adjusting the wavefunction along the steepest descent direction, i.e.~by making a change in $\psi$ proportional to
$ 
\delta\psi=\frac{\delta \ME}{\delta\psi^*}=\frac{1}{nL}\LGP\psi,
$ 
where 
\begin{align}
\LGP=h_\mathrm{sp}+g_s|\psi|^2+\Phidd(\bx) +\gammaQF|\psi|^3,
\end{align}
is the eGPE operator.
Stationary states of the energy functional will satisfy the time-dependent eGPE equation $\LGP\psi=\mu\psi$, where $\mu$ is the chemical potential.  
Here we  primarily use a gradient flow method (also known as the imaginary time method) to find stationary states. We do not describe this method here, but we follow a similar scheme to that described in Ref.~\cite{Lee2021a} (also see \cite{Bao2010b}). However, after a fixed number of gradient flow steps we use a Newton-Krylov approach \cite{Kelley2003a}. The Newton-Krylov approach is particularly beneficial in accelerating convergence of $\psi$ near the transition region where the landscape of the energy functional can be flat. 

We represent the uc wavefunction using a planewave (or Fourier) spectral representation. This is discretized using an equally spaced 3D grid
 on the rectangular cuboid with dimensions $L_\rho\times L_\rho\times L$ and centered on the origin. Here the transverse discretization  range $L_\rho$ is chosen sufficiently large that the wavefunction amplitude decays to a negligible value well-before the transverse boundary. The number of points in each direction is chosen so that the point spacing is  $\Delta\lesssim0.25\,\mu$m.
 This discretization naturally implements the required periodic boundary conditions along $z$ and allows the various terms in the eGPE operator to be computed with spectral accuracy. The DDI term requires special consideration and we discuss this in the next subsection.

Rather than optimise the wavefunction completely to a stationary state we regularly evaluate how the energy per particle changes with uc length and adjust the uc size appropriately. In practice we do this by evaluating the partial derivatives $\left(\frac{\partial \ME}{\partial L}\right)_\psi$ and $\left(\frac{\partial^2 \ME}{\partial L^2}\right)_\psi$ using central finite differences and with this information we perform a Newton-Raphson update to adjust the uc length $L$ .

Our algorithm thus progresses by executing a $\psi$-optimization stage followed by a uc update. This repeats until the residuals, given by
\begin{align}
\mathrm{resid}_\psi&\equiv\max_{\bx}\left|\LGP\psi-\mu\psi\right|,\\
\mathrm{resid}_L&=\left|\left(\frac{\partial \ME}{\partial L}\right)_\psi\right|,
\end{align}
both decrease to values below the termination criteria.\footnote{
These residuals are dependent on the choice of units. Here we use harmonic oscillator units defined by the strongest transverse frequency.
Note that we evaluate the chemical potential as $\mu\equiv\intuc\,\psi^*\LGP\psi/nL$. 
We typically require $\mathrm{resid}_\psi\sim10^{-8}$ and $\mathrm{resid}_L\sim10^{-5}$ for the algorithm to terminate.}

\subsection{Truncated DDI kernel}
The dipolar interactions are conveniently evaluated using the convolution theorem as 
\begin{align}
\Phidd(\bx) = \MF^{-1}\{\UDt(\bk)\MF\{|\psi(\bx)|^2\}\},\label{Phidd2}
\end{align}
 where $\MF$ and $\MF^{-1}$ are the forward and inverse Fourier transforms, respectively.
 Here we have also introduced the $k$-space interaction kernel
  \begin{align}
 \UDt(\bk) =  \gdd\biggl(3\frac{k_y^2}{k^2}-1\biggr),\label{UddBare}
 \end{align} 
 which is the Fourier transform of Eq.~(\ref{Udd}).
This naturally ensures that the periodic repetitions of the density along $z$ are included in the evaluation of $ \Phidd$ [see Eq.~(\ref{PhiDD})]. However, because the transverse directions also have a finite range ($L_\rho$), unphysical periodic copies of the uc density in the transverse plane also contribute to Eq.~(\ref{Phidd2}). The effect of these copies decay as $1/L_\rho^3$, and would require an impractically large transverse spatial grid to reduce their effect to an acceptable level; instead we use a truncated interaction kernel.  This involves limiting the spatial extent of the DDI to a defined spatial region. For the infinite tube trap we use the following truncation:
 \begin{align}
     \UD^R(\br)=\begin{cases}
        \UD(\br), & \rho<R\\
                        0,       & \text{otherwise} 
                    \end{cases}
\end{align}
where  $\rho=\sqrt{x^2+y^2}$ is the radial coordinate. The radial spatial limit ensures that the DDI does not allow (unphysical) transverse copies to interact.
To apply this truncated interaction we need to obtain an accurate Fourier transformed form for use in Eq.~(\ref{Phidd2}). 
 To do this we  express the Fourier transformed kernel as
\begin{align}
  \UDt^R(\bk) &= \UDt(\bk) - \UDt^{R'}(\bk),\label{UdtCut}
  \end{align}
  where ${R'}$ denotes the complementary truncated region $\{\rho\ge R\}$.
  We can evaluate this term as
  \begin{align}
 &\UDt^{R'}(\bk) = \int_R^\infty \rho d\rho \int_{-\infty}^\infty dz \int_0^{2\pi} d\phi \, e^{-i\bk\cdot\br} \UD(\br),\\
  &= 3\gdd [-\tfrac12 l_0(k_\rho R,|k_z|R) + (k_x^2/k_\rho^2-\tfrac12) l_2(k_\rho R,|k_z|R)],
\end{align}
where $\bk\cdot\br = k_\rho \rho \cos(\phi-k_\phi) + k_zz$, and 
using 6.521(2,4) of \cite{Gradshteyn2007a}, for $u>0$, $v>0$, $n\ge0$:
\begin{align}
    l_n(u, v) &\equiv  v^2\int_1^\infty t dt\, K_n(vt) J_n(ut), \\
    &= \frac{ v^2}{u^2+v^2}[v J_n(u) K_{n+1}(v) - u J_{n+1}(u) K_n(v)]. 
\end{align}

 To benchmark this result and demonstrate the usefulness of this cutoff potential we have calculated the dipolar energy per particle,
 \begin{align}
     \ME_D = \frac1{2nL}\intuc\, \Phidd(\bx)|\psi_\mathrm{tr}(\bx)|^2,\label{EDDI}
 \end{align}
with the trial wavefunction,
\begin{align}
\psi_\mathrm{tr}(\bx) = \frac{\sqrt{n}}{\sqrt{\pi}l_{x}l_{y}}e^{-(x^{2}/l_{x}^{2} + y^{2}/l_{y}^{2})/2}.
\end{align}
 We evaluate  Eq.~\eqref{EDDI} numerically, and compare to the exact value
 \begin{align}
 \ME_D^\mathrm{ex} = \frac{n\gdd}{4\pi} \frac{2l_{x}-l_{y}}{l_{x}l_{y}(l_{x}+l_{y})}.
 \end{align}

 \begin{figure}
 \centering
 \includegraphics[trim = 100 250 120 250, width=3in]{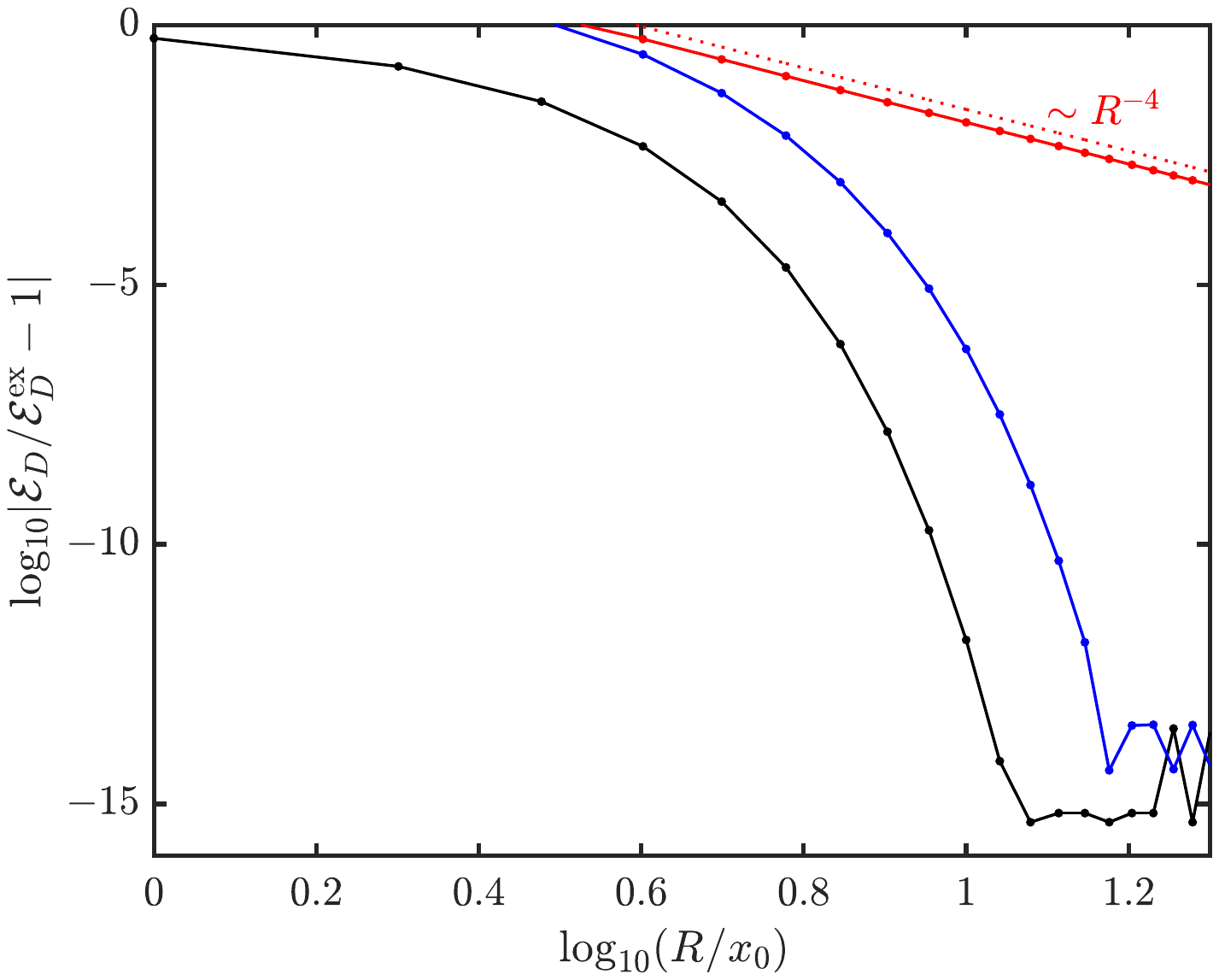}
 \caption{Plot of the relative error in the DDI energy computed using the bare interaction kernel Eq.~\eqref{UddBare} (red) and the cutoff kernel in Eq.~\eqref{UdtCut}  (blue) using $\psi_\mathrm{tr}$ with $l_{x} = \sqrt{2}x_{0}$ and $l_{y}=x_{0}/\sqrt{2}$. The black line shows the relative error in the integration of the density. The grid point density has been fixed to four per unit length.}\label{DDIcomp}
 \end{figure} 

In Fig.~\ref{DDIcomp} we compare the error in calculating the DDI energy with and without a cutoff potential. This shows that in the absence of a cutoff there is a slow decay of the relative error in the energy of the DDI as the size of the grid is increased which demonstrates the inaccuracy that can arise due to improper treatment of the DDI. The relative error decays algebraically $\sim R^{-4}$ when there is no cutoff potential. However, once a cutoff is introduced, the error in this calculation falls off rapidly, converging to machine precision once the state is well represented on the grid.
  
  The singular nature of the DDI potential has limited the development of analytic truncated kernels to two particular cases (see \cite{Ronen2006a}), and our result developed here represents  a new contribution. For other specific geometries the truncated kernels can be obtained numerically \cite{Lu2010a,Tang2017a}, but these are not suitable for the tube situation we consider here. In our algorithm the numerical grid frequently changes (from the $L$ optimisation). This requires constant recomputing of the truncated kernel, making the analytical expression especially well-suited to our problem.

\subsection{Characterising the ground states}

To quantify the translational symmetry that is broken due to density modulation along $z$, we define the density contrast 
\begin{align}
\MC = \frac{n_\mathrm{max} - n_\mathrm{min}}{n_\mathrm{max} + n_\mathrm{min}},
\end{align}
 where $n_\mathrm{max} $ and $n_\mathrm{min}$ are the maximum and minimum of the density $|\psi|^2$ on the $z$-axis. For a uniform state, $\MC=0$, and for a crystalline state $\MC>0$. For $\MC=1$  the density goes to zero on the $z$ axis.

\begin{figure}
\centering
\includegraphics[trim=90 220 90 220,width=3.5in]{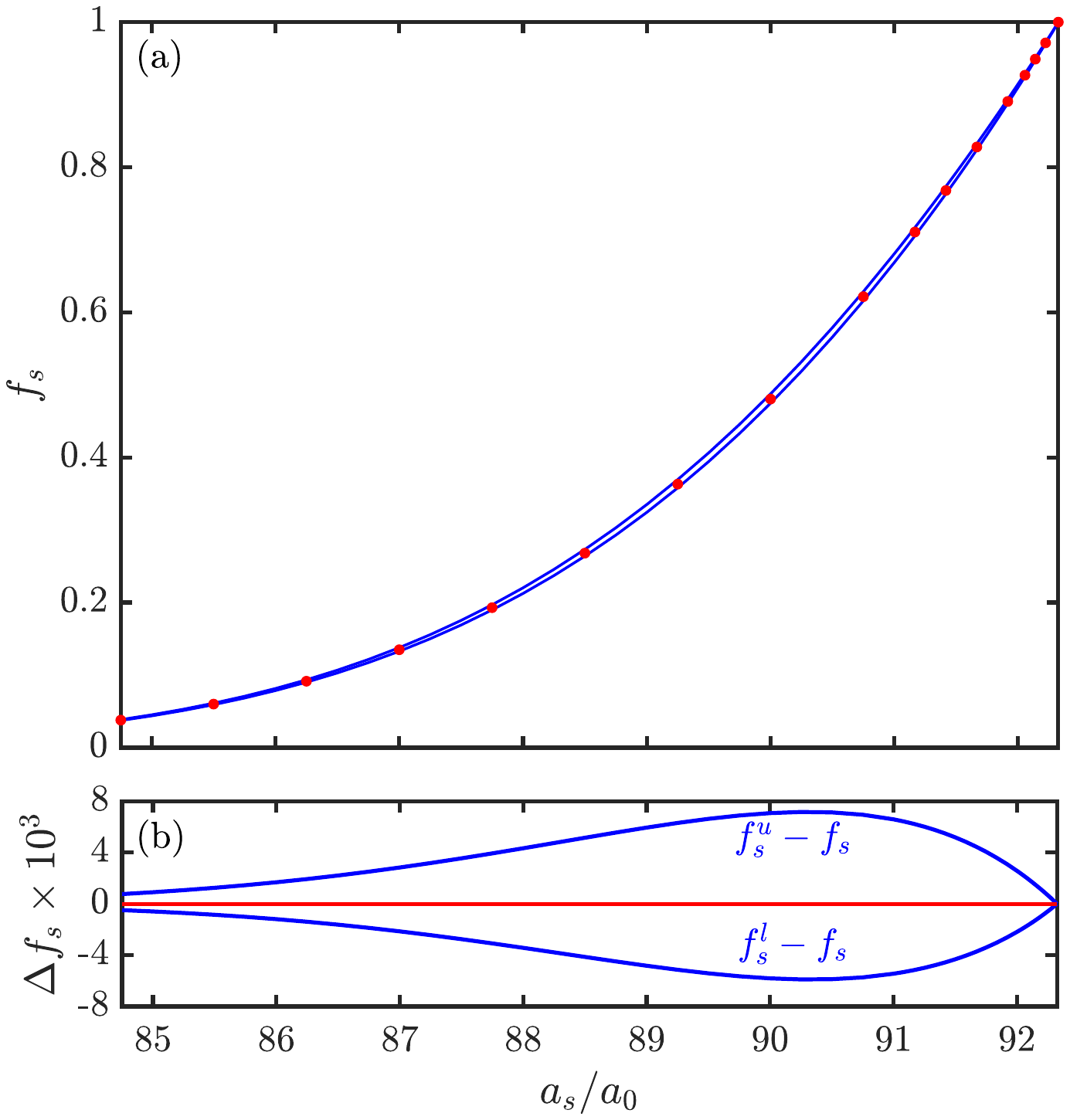}
\caption{(a) Example of the superfluid fraction calculated by nonclassical translational inertia [Eq.~\eqref{fsNCTI}] (red dots), and the upper and lower bounds (blue lines) given by Eqs.~\eqref{sfflegget} and  \eqref{sffleggetLower}, respectively. The difference between the results is barely visible.  (b) The differences $f_s^u-f_s$ and $f_s^l-f_s$ (blue lines). Other parameters: $^{164}$Dy Bose gas with $n=2500/\mu$m and $\omega_{x,y}/2\pi = 150\,$Hz. }\label{Bounds}
\end{figure}

Second, we are interested in quantifying the superfluid fraction of the system. Superfluidity of a Bose-Einstein condensate is associated with a broken global $U(1)$ symmetry. Notably, as pointed out by Leggett \cite{Leggett1998a}  if the translational invariance of the Hamiltonian is broken by the ground state, then the $T=0$ superfluid faction of a Bose gas can be reduced from unity. Thus as the we expect a reduction in the superfluid as the condensate develops crystalline structure. We can most directly quantify the superfluid fraction by computing the nonclassical translational inertia (see \cite{Pomeau1994a,Leggett1998a,Sepulveda2008a}) as
\begin{align}
f_s=1-\lim_{v_z\to0}\frac{\langle P_z\rangle}{nLmv_z},\label{fsNCTI}
\end{align}
where $\langle P_z\rangle=-i\hbar\intuc\,\psi^*\frac{\partial}{\partial z}\psi$ is the expectation of the $z$-component of momentum in a frame moving with uniform velocity $v_z$. 
In Refs.~\cite{Leggett1970a,Leggett1998a} Leggett developed bounds for the superfluid fraction such that $f_s^l\leq f_s \leq f_s^u$, where
\begin{align}
 f_s^u\equiv\frac{L}{n}\left[\int_\mathrm{uc} \frac{dz}{\int d\bm{\rho}\, |\psi|^{2}}\right]^{-1},\label{sfflegget}
\end{align}
and,
\begin{align}
 f_s^l\equiv\frac{L}{n}\int d\brho\,\left[\int_\mathrm{uc} \frac{dz}{ |\psi|^{2}}\right]^{-1},\label{sffleggetLower}
\end{align}
are the upper and lower bounds, respectively.
These expressions can be directly computed from the ground state solution avoiding the need for additional calculations in a moving frame.
While the measures in Eq.~\eqref{sfflegget} and Eq.~\eqref{sffleggetLower}  are  bounds of the superfluid fraction, for the infinite tube dipolar gas we have confirmed that $f_s$, $ f_s^u$ and, $ f_s^l$ are all in good agreement.\footnote{In 1D cases \cite{Leggett1998a,Sepulveda2008a}) or situations where the wavefunction is separable (e.g.~reduced theory presented in~Sec.~\ref{Sec:Reducedthry}) we have $f_s=f_s^u= f_s^l$.} Notably, for the results in Fig.~\ref{Bounds} the maximum difference from the bounds to  $f_s$ is $\sim0.7\%$. The moving frame measure is more difficult to apply near the transition\footnote{Equation (\ref{fsNCTI}) is evaluated using the momentum expectation of the state in a slowly moving frame. For parameters very close to the transition it can be difficult to obtain a modulated solution in the moving frame as the additional kinetic energy can cause it to convert to a uniform to state. } and for this reason the superfluid fraction results we present here are evaluated using Eq.~(\ref{sfflegget}).

\subsection{Reduced 3D theory}\label{Sec:Reducedthry}
 The reduced 3D theory was introduced in Ref.~\cite{Blakie2020b}, and we briefly review it here as it is used to compare against the full 3D results for the phase diagram. The basis of this theory is to decompose the 3D field as $\psi(\mathbf{x})=\phi(z)\chi_\mathrm{var}(\brho)$ where $\chi_\mathrm{var}(\brho)=\tfrac{1}{\sqrt{\pi}l} {e^{-(\eta x^2+y^2/\eta)/2l^2}}$ is a two-dimensional Gaussian function with variational parameters $\{l,\eta\}$, and $\phi(z)$ describes the axial field with $\int_{\mathrm{uc}} dz\,|\phi|^2= nL$ and periodic boundary conditions. The energy per particle of the reduced theory is given by
\begin{align}
\ME_\mathrm{red}=\ME_\perp+\int_{\mathrm{uc}}\!\frac{dz}{ nL}\,\phi^*
\!\left(\!-\frac{\hbar^2}{2m}\frac{d^2}{dz^2} +\frac{1}{2}\Phi +\frac{4\gammaQF|\phi|^3}{25\pi^{3/2}l^3}\right)\!\phi\label{eGPE_Einf}, 
\end{align}
where
\begin{align}
\ME_\perp = \frac{\hbar^2}{4ml^2}\left(\eta+\frac{1}{\eta}\right)+\frac{ml^2}{4}\left(\frac{\omega_x^2}{\eta}+\omega_y^2\eta\right),
\end{align}
is the single-particle energy of the transverse degrees of freedom. The two-body interactions are described by the effective potential  
\begin{align}
\Phi (z)=\MF_z^{-1}\left\{ \Ut(k_z ) \MF_z\{|\phi|^2\}\right\},
\end{align}
  with $\MF_z$ being the 1D Fourier transform, and where
  \begin{align}
      \Ut(k_z)& = \frac{g_s}{2\pi l^2}  + \frac{\gdd}{2\pi l^2} \!\left\{\!\frac{3[ Qe^{Q}\Ei(-Q)+1]}{1+\eta}-1\!\right\},\label{Ueta} 
\end{align} 
with  $\Ei$ being the exponential integral, and $Q\equiv \tfrac{1}{2}\sqrt{\eta}k_z^2l^2$ (see \cite{Blakie2020a}).  
   To obtain stationary solutions of the reduced theory we vary $\{l,\eta,L\}$ and $\psi(z)$ to find local minima of (\ref{eGPE_Einf}).  This theory is rather efficient to solve and was used to construct a  phase diagram for the infinite tube dipolar gas  in Ref.~\cite{Blakie2020b}.

\begin{figure}[htp!]
\begin{centering}
    \includegraphics[trim=0 0 0 450,clip=true,width=3.5in]{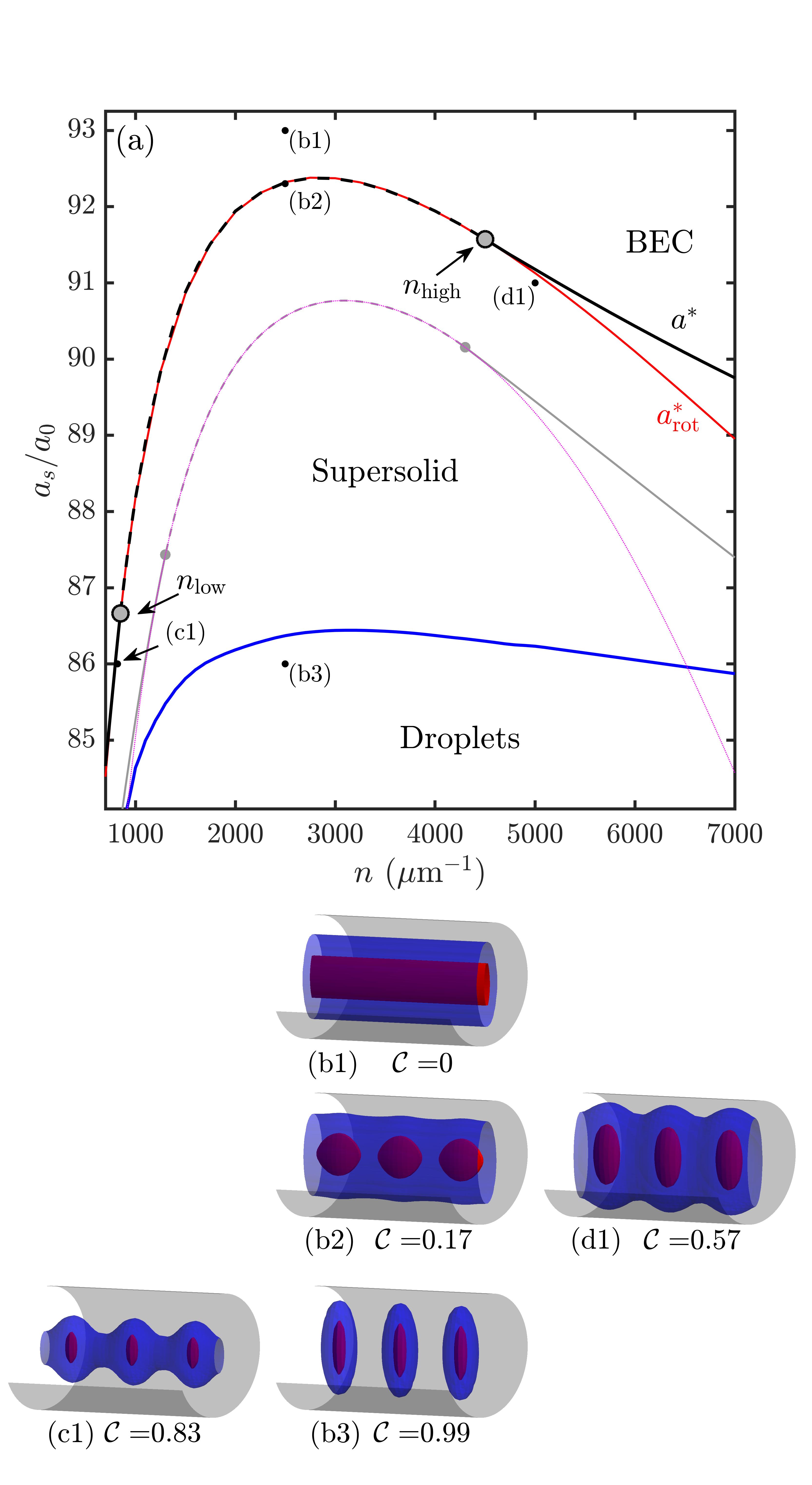}
	\caption{(a) Phase diagram for a $^{164}$Dy Bose gas confined by a radially symmetric infinite tube trap with $\omega_{x,y}/2\pi = 150\,$Hz. The black line indicates the crystallization transition boundary $a^*$. This line is solid where the transition is discontinuous and is dashed where it is continuous. The circles demarcate the start and the end of the continuous region at $n_\mathrm{low}$ and $n_\mathrm{high}$, respectively.  The parameters where a roton excitation in the uniform state softens to zero energy are shown with the red line. The crystallization transition boundary of the reduced 3D theory is shown in grey, and the roton boundary from that theory in magenta. (b1)-(b3),(c1)-(d1) Example ground states for parameters indicated in subplot (a). Isodensity surfaces of $|\psi|^2$ (repeated over 3 unit cells for the modulated cases) taken at $75\%$ (red) and $1\%$ (blue) of the peak density. The grey isosurface indicates a cutaway isoenergy surface of the trapping potential.}\label{fig:PD}
\end{centering}
\end{figure}

\section{Results}\label{Sec:Results}
\subsection{Phase diagram}

The results presented here are for a Bose gas of $^{164}$Dy atoms with $\add = 130.8a_0$, confined to a radially symmetric trap with $\omega_{x,y}/2\pi=150\,$Hz. This system is similar to that considered for the main results of Ref.~\cite{Blakie2020b} calculated using the reduced theory.\footnote{For the theory presented here we have  used $\mathcal{Q}_5$ in the definition of the quantum fluctuation term rather than the quadratic approximation used in Ref.~\cite{Blakie2020b}. This gives rise to a small shift in the predictions.}  Figure~\ref{fig:PD}(a) is the phase diagram  obtained by minimising the energy per particle \eqref{ucEfunc} as $a_s$ and $n$ are varied, and is the main result of this paper.
We identify three different phases using the ground state properties. (i) The uniform \textit{Bose-Einstein condensate} (BEC)  is translationally invariant along $z$, with $\MC=0$ and $f_s=1$ [see Fig.~\ref{fig:PD}(b1)]; (ii) The \textit{supersolid state} is a crystalline state with $\MC>0$, yet retaining a finite superfluid fraction $f_s$ [see Figs.~\ref{fig:PD}(b2), (c1), and  (d1)]; (iii) The \textit{insulating droplet state} is a crystalline state with tightly-bound and separated droplets  [see Fig.~\ref{fig:PD}(b3)]. For the insulating droplet state the contrast of the density modulation along $z$ is essentially complete ($\MC\approx1$) and the superfluid fraction is negligible\footnote{Here we  follow Ref.~\cite{Blakie2020b} and  set  a superfluid fraction of $f_s^{\min}=0.1$  to distinguish between the supersolid and insulating droplet states.} $f_s\approx0$. We do not concern ourselves with the nature of the transition or crossover from supersolid to insulating droplets. A proper description of this is beyond the eGPE theory, and requires a theory capable of capturing correlations between droplets needed to drive the insulating transition.

\begin{figure}
\begin{centering}
	\includegraphics[trim=90 220 90 220,width = 3.5in]{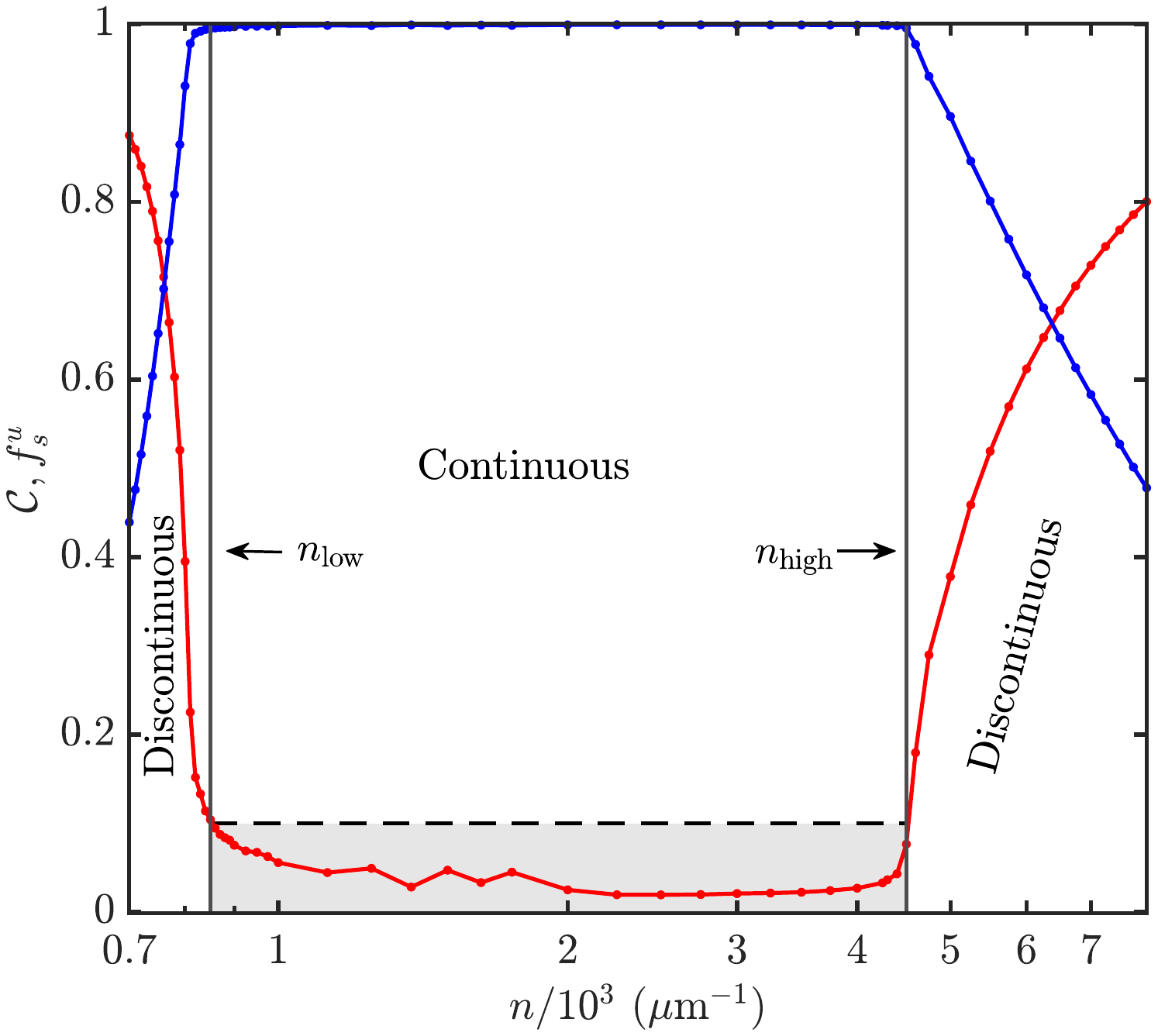}
	\caption{Superfluid fraction (blue) and density contrast (red) as a function of the average linear density of the crystalline state along the transition boundary $a^*$ [see Fig.~\ref{fig:PD}]. 
	The vertical lines identify the low density ($n_\mathrm{low}$) and high-density ($n_\mathrm{high}$) end points of the region in which the transition is continuous.
	The grey-shaded region indicates contrast values which we take to be zero (see text). Other parameters as in   Fig.~\ref{fig:PD}. }\label{fig:TransitionLine}
\end{centering}
\end{figure}

\begin{figure}[htb!]
\begin{centering}
	\includegraphics[trim=90 220 60 220,width = 3.5in]{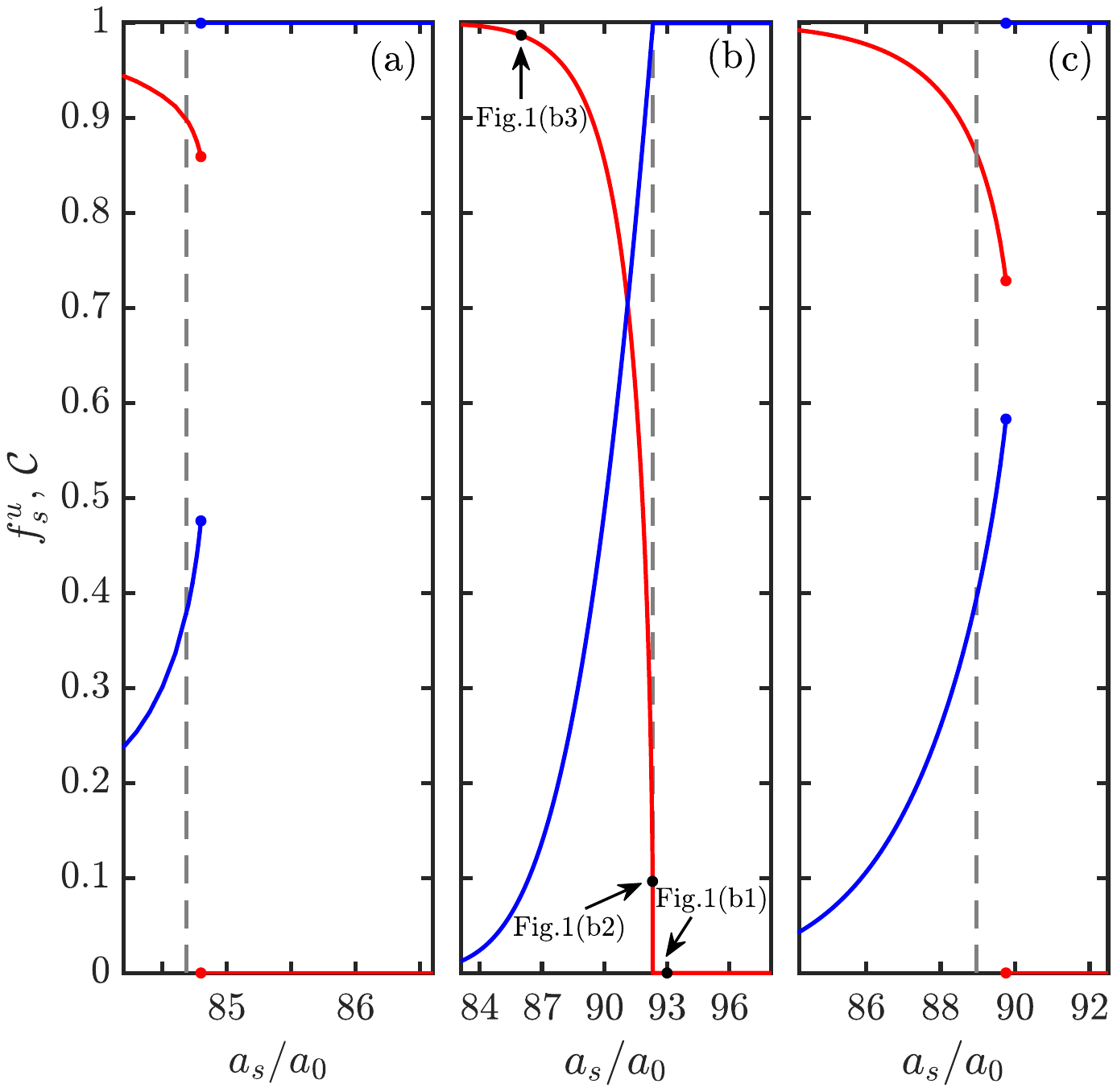}
	\caption{Behavior of the ground state density contrast (red) and superfluid fraction (blue) as $a_s$ varies across the transition for three different cases: (a) $n=710/\mu$m, (b) $n=2500/\mu$m, and (c) $n=7000/\mu$m. The vertical grey dashed lines indicate $\arot$ for each case. Other parameters as in Fig.~\ref{fig:PD}.
 }\label{fig:pd_cases}
\end{centering}
\end{figure}

We define $a^*$ as the value of $a_s$ at which the ground state transitions  from a uniform BEC to having crystalline order (i.e.~$\MC=0$ to $\MC\ne0$) for a system with average linear density $n$. Over the  range  of $n$  values shown in Fig.~\ref{fig:PD}(a) the transition occurs directly from the uniform BEC  state into a supersolid state, with the insulating droplet state emerging at lower values of $a_s$.
 We observe that $a^*$ initially increases with $n$ as the role of interactions increases relative to kinetic energy, but then $a^*$ decreases for $n\gtrsim2.8\times10^3/\mu$m, as  quantum fluctuation effects become stronger.\footnote{The quantum fluctuation term acts as a local repulsive interaction, thus at high $n$ the $a_s$ value needs to be reduced further for the DDIs to overcome the local repulsive interactions and drive the system to crystallize.} 
Our main interest here is the nature of the BEC to crystalline transition. 
At each density $n$ we have obtained the crystalline ground state solution at the transition point $a^*$, where it is degenerate with the uniform BEC state. The contrast and superfluid fraction of these states are shown in Fig.~\ref{fig:TransitionLine}, revealing that for an intermediate density range $n\in[n_\mathrm{low},n_\mathrm{high}]$ with $n_\mathrm{low}\approx0.8\times10^3/\mu$m  and $n_\mathrm{high}\approx4.5\times10^3/\mu$m, the transition is continuous. That is, in this range $\MC\to0$ and $f_s\to1$ as $a_s\to a^*$ from below. The end points of this continuous transition line are indicated by two circles in Fig.~\ref{fig:PD}(a).

In Fig.~\ref{fig:pd_cases} we examine the variation of the superfluid fraction and the density contrast of the ground state as a function of $a_s$.   We present a low density discontinuous case [Fig.~\ref{fig:pd_cases}(a)], an intermediate density continuous case [Fig.~\ref{fig:pd_cases}(b)], and a high density discontinuous case [Fig.~\ref{fig:pd_cases}(c)]. These results also show that in the discontinuous transition the superfluid fraction of the crystalline state is significantly less than unity, even for $a_s$ values very close to the transition. In contrast, in the vicinity of the continuous transition the superfluid fraction of the crystalline state tends to be close to unity for an appreciable range of $a_s$ values below $a^*$.

Since crystalline order causes a reduction in the superfluid fraction we can use either $f_s$  or $\MC$ to quantify the crystalline transition (e.g.~cf.~Refs.~\cite{Roccuzzo2019a,Blakie2020b,Biagioni2022a}). The density contrast $\MC$ is most frequently used in application to the experimental regime  with a cigar shaped harmonic trap. Our results show that immediately below the continuous transition point $\MC$ charges more rapidly with $a_s$ than $f_s$ [see Fig.~\ref{fig:pd_cases}(b)]. This can be understood by a simplified analytic model of the weakly modulated state developed in Ref.~\cite{Blakie2020b}, which shows that in the vicinity of the continuous transition for small $\MC$, the reduction in superfluid fraction is second order in $\MC$ \cite{Chomaz2020a}
\begin{align}
f_s = 1-\frac{1}{2}\MC^2 + O(\MC^4).\label{fsCfromCM}
\end{align}

If the transition is continuous or almost continuous then, near the transition, (weakly) crystalline and uniform states have very similar energy, which makes precisely determining $a^*$ challenging. For the results presented in Fig.~\ref{fig:TransitionLine}, we have taken cases with $\MC<0.1$ to be equivalent to $\MC=0$ (i.e.~results in the grey shaded region), and identified these as being the continuous transition regime. We find $a^*$ to an accuracy of $\sim10^{-4}\,a_0$, which requires resolving a difference in the relative energy per particle of crystalline and uniform states to $\sim10^{-7}$. In contrast, the difference in the numerically obtained $f_s$ from unity at $a^*$ is almost too small to notice in Fig.~\ref{fig:TransitionLine}, i.e.~the scatter in $f_s$ is less than 1\% [consistent with Eq.~\eqref{fsCfromCM}].

For comparison, we also indicate the phase transition from the reduced theory in Fig.~\ref{fig:PD}(a). The transition boundary $a^*$ predicted by the reduced theory is approximately $1\,a_0$ to $2\,a_0$ lower than that of the full eGPE calculation. This shift arises from the variational treatment of the transverse degrees of freedom in the reduced theory, and is consistent with other such comparisons of the reduced theory to full 3D calculations (e.g.~see the shifts in scattering lengths for the occurrence of rotons noted in Fig.~5 of \cite{Blakie2020a}).

\subsection{Roton softening and the continuous transition}

\begin{figure}
\begin{centering}
    \includegraphics[trim=90 220 90 220,width = 3.3in]{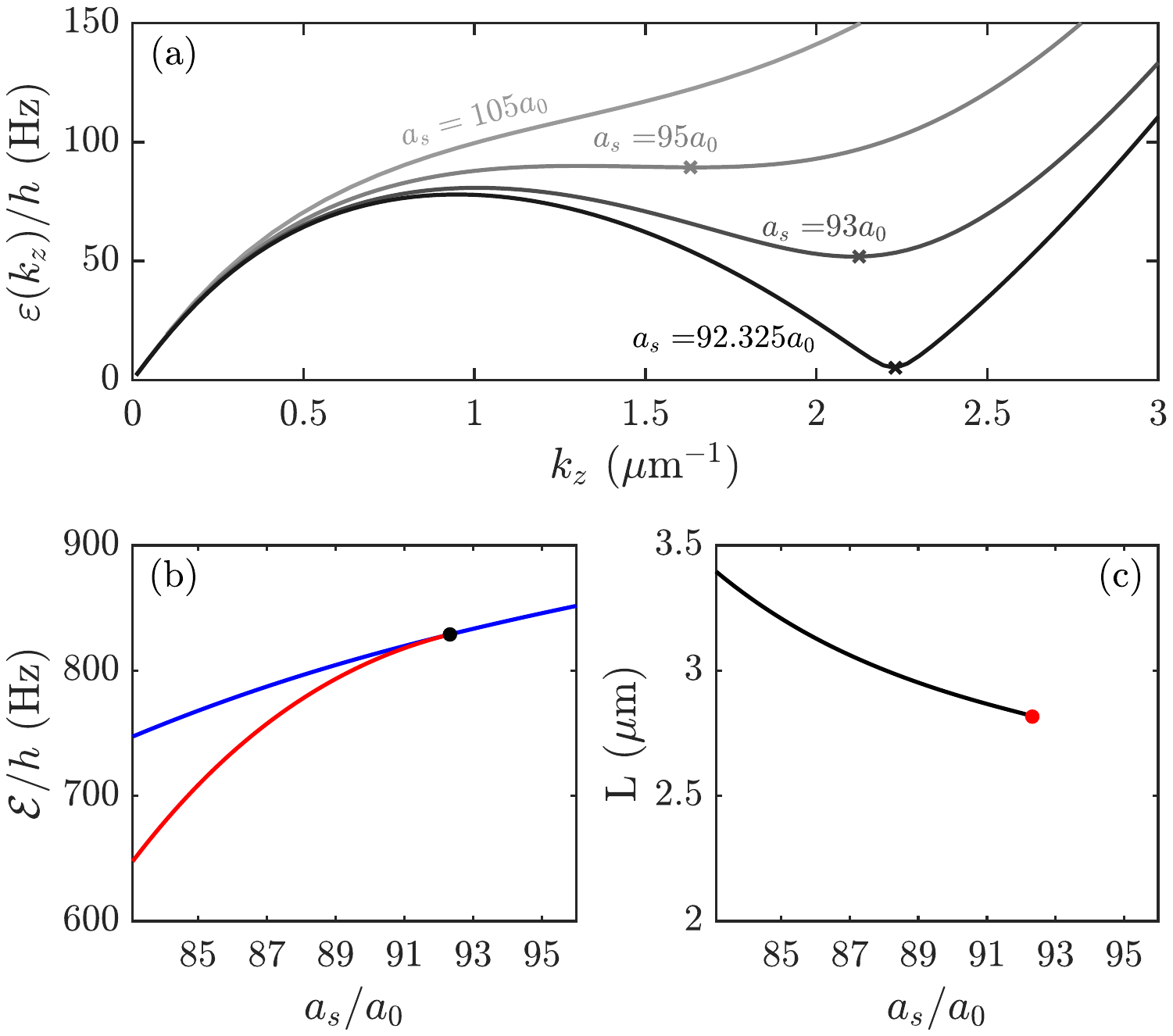}
	\caption{ (a) Dispersion relation of the lowest Bogoliubov excitation band for a uniform BEC state for $a_s$ values as indicated. A roton excitation manifests as a local minimum in this dispersion relation (marked with $\times$). The roton softens to zero energy as $a_s\to \arot$ from above. (b) Energy per particle of the uniform (blue line) and modulated (red line) states; the black dot indicates where the energies cross.  (c) Length $L$ of the  uc  (undefined out of the crystalline phase). The red dot has coordinates $(\arot,\lambda_\mathrm{rot}$) where $\lambda_\mathrm{rot}=2\pi/k_\mathrm{rot}$
    is the roton wavelength and $k_\mathrm{rot}>0$ is the wavevector where the dispersion touches zero. Results for $n=2500/\mu$m and other parameters as in  Fig.~\ref{fig:PD}.\label{fig:excitation}}
\end{centering}
\end{figure}

We determined the collective excitations of the uniform BEC state by solving the Bogoliubov-de Gennes equations (see Ref.~\cite{Blakie2020a} for details). As the uniform groundstate is translationally invariant, the  excitations can be characterized by the $z$-component of momentum ($\hbar k_z$), and occur as a set of bands with different transverse excitation. 
In Fig.~\ref{fig:excitation}(a) we show the results for the lowest excitation band for a system with $n=2500/\mu$m and at various $a_s$ values.
 At the highest value shown ($a_s = 105a_0$) the spectrum is a monotonically increasing function of $k_{z}$. As $a_s$ is lowered, a local minimum develops at a non-zero wavevector, a roton-like mode which arises in dipolar BECs from the interplay of the DDIs and confinement \cite{Santos2003a,Chomaz2018a,Petter2019a}. 
The minimum energy of the roton decreases with decreasing $a_s$, until the energy softens to zero energy. We identify the value of scattering length when this occurs as $\arot$, and the corresponding $k_z$ value of the zero-energy mode as the roton wavevector $k_\mathrm{rot}$. For $a_s<\arot$  the roton is dynamically unstable, indicating that the uniform BEC state is no longer the ground state [see Fig.~\ref{fig:excitation}(b)]. For the density considered in these results  $\arot=a^*$ [cf.~Fig.~\ref{fig:pd_cases}(b)] and the roton softening marks the critical point at which the crystalline order continuously emerges. In Fig.~\ref{fig:excitation}(c) we show that the uc length $L$ of the crystalline ground state corresponds to the roton wavelength at the critical point.

We indicate the roton instability as a function of density in the phase diagram, Fig.~\ref{fig:PD}(a). We see that  where a continuous transition is found, the roton instability coincides with the transition, i.e.~$a^*=\arot$ for $n\in [n_\mathrm{low},n_\mathrm{high}]$. Outside of this region $\arot<a^*$, such that the uniform BEC state is energetically unstable for $a_s<a^*$, but not dynamically unstable until $a_s<\arot$. 

Ref.~\cite{Roccuzzo2019a} considers a finite tube with periodic boundary conditions and a fixed unit cell size, and finds a discontinuous $\sim10\%$ jump in $f_s$ at the crystallization transition. We have performed calculations for those parameters ($^{166}$Er Bose gas with $n=3.78\times10^3/\mu$m and $\omega_{x,y}/2\pi=600\,$Hz), but for an infinite tube with the unit cell size chosen to minimize energy. We find that the transition to the crystalline transition is continuous and occurs with $a^*=\arot$ [similar to Fig.~\ref{fig:excitation}(a) and \ref{fig:pd_cases}(b)]. The fixed system length (e.g.,~not being an integer multiple of $\lambda_\mathrm{rot}$)   and the absence of a truncated DDI potential for the calculations in Ref.~\cite{Roccuzzo2019a} may have contributed to the prediction of a discontinuous transition.

\subsection{Discontinuous phase transition}

\begin{figure}
\begin{centering}
	\includegraphics[trim=70 230 70 230,width = 3.5in]{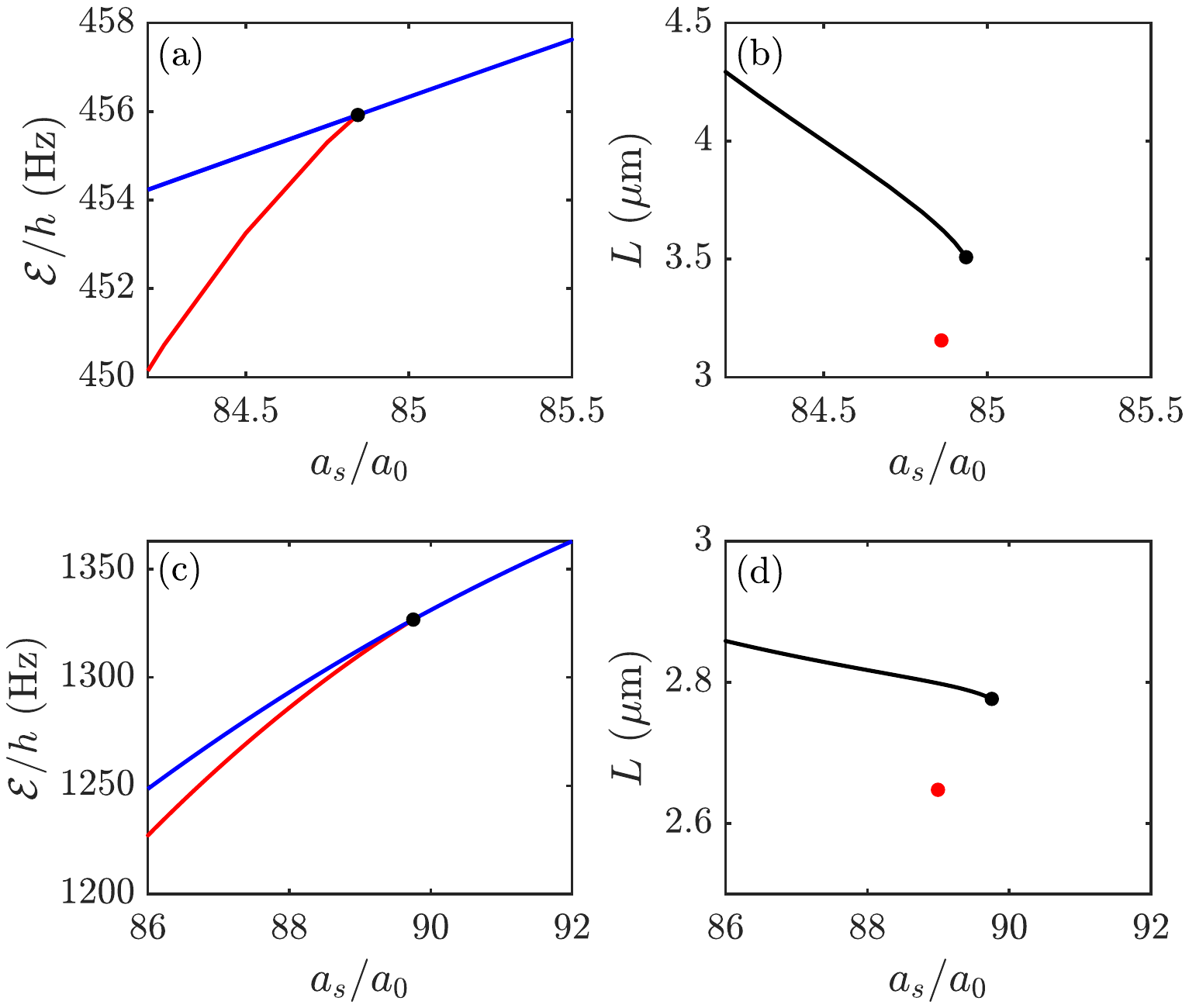}
    \caption{ (a,c) Energy per particle of modulated state (red) and uniform state (blue); the black dot indicates where the energies cross.  (b,d) Length of the uc for the crystalline state.  The red dot has coordinates $(\arot,\lambda_\mathrm{rot}$).   (a,b) $n=710/\mu$m and (c,d) $n=7000/\mu$m, with other parameters as in Fig.~\ref{fig:PD}. 
     }\label{fig:EDUL}
\end{centering}
\end{figure}

In Fig.~\ref{fig:EDUL} we examine the behavior of the transition for low- and high-density discontinuous cases. Here the energy per particle of the uniform and crystalline states cross [difficult to see in Figs.~\ref{fig:EDUL} (a) and (c)]. We are unable to follow the crystalline branch to values of $a_s$ much higher than $a^*$ because the gradient flow algorithm is an energy minimisation scheme and causes the crystalline state to jump to the uniform branch. In contrast, we are able to solve for the uniform solution for $a_s<a^*$ using the ansatz $\psi=\sqrt{n}\chi(\brho)$, which does not allow any modulation to develop along $z$. The uc length comparison in Figs.~\ref{fig:EDUL}(a) and (c) also shows that the crystalline states have $L>\lambda_\mathrm{rot}$ in the region of the discontinuous transition.
The high density discontinuous transition occurs in a regime where the quantum fluctuation effects are relatively strong and has been identified as a fluctuation induced first order transition in the finite system studied in Ref.~\cite{Biagioni2022a}.

\newpage
 
\section{Conclusions}\label{Sec:Conlcusion}

In this paper we have developed and applied a numerical method to study the ground states of a dipolar Bose gas in an infinite tube potential. We perform these calculations within an optimized unit cell, utilizing an analytic result for a truncated DDI kernel.   Using our method, we have explored the  BEC to crystalline transition as a function of the average linear density for a system with isotropic transverse trapping. We find that a continuous transition to the crystalline state occurs for intermediate densities, and that in the vicinity of the continuous transition, the crystalline state has a significant superfluid fraction, and thus may be a useful regime to study supersolidity. These results demonstrate that the reduced 3D theory developed in Ref.~\cite{Blakie2020b}   provides a qualitatively accurate description of the transitions in this system, however the transition lines of the reduced 3D theory are shifted to lower values of $a_s$. Our results verify that the continuous transition occurs coincident with the softening of a roton excitation in the uniform BEC state, which initiates the crystallization at the roton wavelength. For densities where the transition is discontinuous, the roton softens at a value of $a_s$ below  $a^*$. This behavior may allow hysteresis to occur such that, as $a_s$ is ramped across the transition, the BEC could persist to values of $a_s$ below $a^*$. Similarly, the crystalline state could persist to values of $a_s$ above $a^*$.

In future work we will study the behavior of the transition in the low density regime in more detail. It is of interest to understand if the transition to the crystalline state in this regime occurs directly into a well-separated insulating droplet state or some other arrangement. Another issue is to understand the role of the transverse confinement, especially when it is anisotropic.
 A deeper understanding of these effects will provide insight into the nature of the transition in the finite  system (e.g.~3D cigar trap, see \cite{Biagioni2022a,Alana2022a}), and will provide a point of comparison to work looking at crystallization and supersolids in 2D  (e.g.~see \cite{Lu2015a,Zhang2019a,Baillie2018a,Zhang2021a,Norcia2021a,Schmidt2021a,Hertkorn2021b,Poli2021a,Hertkorn2021b}).
 
 \section*{Acknowledgment}
We acknowledge useful discussions with A.-C. Lee, F.~Ferlaino, L.~Chomaz, S.~Roccuzzo and R.~Bisset and support from the Marsden Fund of the Royal Society of New Zealand.


%

\end{document}